# Electromagnetic Field Behavior in Isotropic Negative Phase Velocity/Negative Refractive Index Guided Wave Structures Compatible with Millimeter-Wave Monolithic Integrated Circuits

Clifford M. Krowne and Maurice Daniel*

Microwave Technology Branch, Electronics Science & Technology Division, Naval Research Laboratory, Washington, D.C. 20375-5347

*DCS Corporation, 1330 Braddock Place, Alexandria, VA 22314

**ABSTRACT**

A microstrip configuration has been loaded with an isotropic left-handed medium (LHM) substrate and studied regarding its high frequency millimeter wave behavior near 100 GHz. This has been accomplished using a full–wave integral equation anisotropic Green's function code configured to run for isotropy. Never before seen electromagnetic field distributions are produced, unlike anything found in normal media devices, using this ab initio solver. These distributions are made in the cross-sectional dimension, with the field propagating in the perpendicular direction. It is discovered that the LHM distributions are so radically different from ordinary media used as a substrate, that completely new electronic devices based upon the new physics becomes a real possibility.

# I. Introduction

Tremendous interest in the last few years has occurred with the experimental realization of macroscopic demonstrations of left–handed media, predicted or at least suggested in the literature several decades ago [1]. Attention has followed on the focusing characteristics and related issues of left–handed media (LHM), with appropriate arrangements to accomplish such behavior, as shown by literature publications [2 – 25]. But no attention has been directed toward what intrinsic left–handed media could do in propagating devices used in integrated circuit configurations. This is not to say that some work has not happened on applications using backward wave production or LHM properties in specialized microwave, devices which rely on reduced dimension negative phase velocity behavior [26] – [34]. (Also see references contained in [35] and [36] for other focusing and backward wave devices.) And much of that work has looked at macroscopic realizations, which may be amenable in the future with current efforts on metamaterials, to advancing microwave integrated circuit component technology utilizing left–handed media.

We are particularly interested here in what new physical properties are the result of using material which is intrinsically left–handed, or also variously referred to in the literature as negative phase velocity material (NPVM or NPV) or negative refractive index material (NRIM or NIM). There may be substantial interest in understanding the effects of left–handed media in guided wave structures since advances in integrated circuit technology, in passive components, control components, and active devices has increasingly been utilizing layers and arrangements of many differing materials. From heterostructures in active devices to complex materials like chiral, ferroelectric and ferromagnetic materials, in passive and control components, this trend has been rising. Efforts on metamaterials is sure to further this trend.

A hint at the remarkably different field distributions has been disclosed recently using LHM substrates in guided wave devices [37]. Dispersion diagram description of the physics is provided in [35], and this diagram shows the effect of the RHM/LHM interface seen in the cross–sectional view on the propagation normal to the cross–section. Bands of pure phase propagation and bands of evanescent propagation occur. Also, negative phase velocity behavior of the LHM interacts with the RHM to generate regions of both ordinary wave propagation as well as backward wave propagation in the negative phase velocity sense relative to the guided wave power flow. In [35] only the low end band is displayed in field distribution plots at 5 GHz. But the plots shown are instructive for the new physics they demonstrate: unusual field line or circulation characteristics for the electric or magnetic fields, startling intensity variation of the fields, counter intuitive charge arrangement on the guiding metal strip, and interesting visual display of opposed Poynting vectors in adjoining RHM and LHM regions for power flow down the device.

Attention to the new possibilities for electronic devices is given in [36], [38] when using LHM/NPV substrates. There, distributions up to 40 GHz are provided, somewhat over the beginning of the millimeter wave frequency band. However, nowhere have we made available the remarkable field distributions found at the higher millimeter wavelengths, and so in this article we would like to show for the first time what the fields look like at nearly $10^{11}$ = 100 GHz (we will actually draw our attention to f = 80 GHz as a starting point). A new technique we have developed of lifting out the lower magnitude fields in order to visualize their directions in arrow distribution plots will be utilized here for the first time (Section V). This is particularly important in distribution plots where the field magnitudes may vary over many orders of magnitude. A number of field distribution plotting methods will be employed in this paper: arrow plots based upon linear representation of the field magnitude, arrow plots for both electric and magnetic fields based upon scaled representation of the field magnitude, line plots showing electric field behavior emanating from the strip and off of it, line plots showing magnetic field behavior circulating around the guiding strip and off of it, and magnitude plots of both the electric and magnetic fields.

Sections II and III provide short discussions of left–handed material properties (Section II) and the Green's function technique (Section III) used to solve the material physics/field problem. Once these preliminaries are out of the way, the eigenvalues (Section IV) and the field distributions (Section V) are determined.

## II. Left – Handed Material Characteristics

It is expected that the left–handed medium's characteristic to alter the electromagnetic field based upon its new properties contained in its tensors describing permittivity and permeability will not only lead to new structures enlisting just this new material, but eventually allow the creation of multi–layered devices containing various substances including left–handed media. Here we report on the new physics associated with left–handed media in guided wave propagating structures which are applicable to microwave and millimeter wave integrated circuits, although the focus here is primarily on the millimeter wave region. Here we address the use of the left–handed media with its general bianisotropic crystalline properties reduced to scalars, that is with anisotropic permittivity $\bar{\bar{\varepsilon}}$ tensor set equal to the isotropic permittivity value $\varepsilon \bar{\bar{1}}$, and anisotropic permeability $\bar{\bar{\mu}}$ tensor set equal to the isotropic permeability value $\mu \bar{\bar{1}}$. Consideration of the anisotropic or bianisotropic crystalline case is examined elsewhere [39]. Suffice it to say here that just as in the case of optical or lower frequency focusing, isotropy is what allows proper organization of all the wave fronts (or rays in the geometric optics limit). But in a guided wave structure, what may be the most critical issue, is the assumption of isotropy to allow arbitrary field contouring or sculpting [40]. Individual unit cell construction and repetitive cells in all directions can lead to isotropy, as well as materials with intrinsic isotropic crystalline properties. The scalar relative permittivity and permeability $\varepsilon$ and $\mu$ seen in the literature have frequency dependence $\varepsilon(\omega)$ and $\mu(\omega)$. Left–handed material is obtained when $\text{Re}[\varepsilon(\omega)] < 0$ and $\text{Re}[\mu(\omega)] < 0$ simultaneously. Whether this is a narrow or wide band phenomenon will not be addressed here, other than to note that there may be both metaobject construction as well as intrinsic material methods to adjust the bandwidth. There is every indication today that these two implementation categories may provide enough design possibilities to make such bandwidth adjustment realistic. So whether one uses nonresonant objects or resonant objects, or microscopic properties of crystals or nanoscale materials, there is no reason to doubt that the frequency region $\Delta\omega$ over which the desired behavior occurs may be viewed as being subject to the choice of the physicist or engineer for some intended use.

So in order to study what the field distributions for a LHM substrate would do in a certain configuration at a particular frequency, we need to set $\text{Re}[\varepsilon(\omega)] = -\varepsilon_r$ and $\text{Re}[\mu(\omega)] = -\mu_r$ where $\varepsilon_r$ = real positive constant and $\mu_r$ = real positive constant. Fundamental mode is sought, which has an eigenvalue even as $\omega \to 0\ (f \to 0)$. This is the simplest problem one can solve for in our inhomogeneous boundary condition problem. Hope for obtaining wideband behavior is now supported by recent results showing that negative refraction can be obtained by using heterostuctures of intrinsic crystals with negligible dispersion [41] – [44] . Narrowest behavior occurs with resonant structures like split ring resonator-rod combinations. Even for these structures, which are characterizable by $\varepsilon(\omega) = 1 - (\omega_p^2 - \omega_0^2)/(\omega^2 - \omega_0^2 + i\omega L_\varepsilon)$ and $\mu(\omega) = 1 - F\omega^2/(\omega^2 - \omega_0^2 + i\omega L_\mu)$ (these forms are widely quoted in the literature, with, for example, the permeability being derivable from [45]), because $\omega_p$, $\omega_0$ and F are subject to the designer's control, one can always, for a desired setting for $\varepsilon$ and $\mu$ at a particular frequency $\omega$, solve the two equations implicit in these depictions for the three unknowns, with a rich multiplicity of solutions. Finally, photonic crystals which provide the negative refraction using ordinary RHM crystals with RHM inclusions, may have bandwidths somewhere between non-dispersive and highly dispersive (as in the resonant structures).

### III. Green's Function for Left – Handed Guided Wave Structure

Green's function for the problem is a self–consistent one for a driving surface current vector Dirac delta function applied at the guiding microstrip metal, $\mathbf{J} = j_x \delta(x - x_0)\hat{x} + j_x \delta(x - x_0)\hat{z}$ with $x_0$ = point on the strip. Figure 1 (a) shows a cross-section of the structure with a right-handed material (RHM/PPV - ordinary material; PPV = positive phase velocity) used for the substrate and Fig. 1 (b) shows a cross-section of the structure with a left-handed material (LHM/NPV). The Green's function is a dyadic, constructed as a $2 \times 2$ array relating tangential x- and z- components of surface current density to tangential electric field components. This Green's function is used to solve for the propagation constant (see [46] for a recent use of this type of Green's function). Determination of the field components is done in a second stage of processing, which in effect creates a large rectangular Green's function array, of size $6 \times 2$, in order to generate all electromagnetic field components, including those in the y – direction normal to the structure layers. The governing equation of the problem can be stated as

$$\frac{d\psi}{dy} = i\omega \mathbf{R} \psi \; ; \; \psi = \begin{bmatrix} E_x & E_z & H_x & H_z \end{bmatrix}^T \qquad (1)$$

where the system $4 \times 4$ matrix $\mathbf{R}$ depends on the Green's function and the physical properties of the materials. This equation gives the tangential transverse field component variation (column vector $\psi$) perpendicular to the surface in the y – direction. Auxiliary equations give the two remaining field components, $E_y$ and $H_y$.

The self–consistent problem is solved by expanding the surface currents on the guiding microstrip metal in infinite expansions $J_x = \sum_{i=1}^{n_x} a_{xi} j_{xi}(x)$ and $J_z = \sum_{i=1}^{n_z} a_{zi} j_{zi}(x)$ and then requiring the determinant of the resulting system of equations to be zero. At this step of the problem, only the surface current basis functions $j_{xi}(x)$ and $j_{zi}(x)$ need be provided and the complex propagation constant $\gamma = \alpha + j\beta$ is returned by the computer code. Of course, the summation limits $n_x$ and $n_z$ must be truncated at an appropriate value when convergence is acceptable.

Acquisition of the electromagnetic fields necessitates obtaining the basis function expansion coefficients $a_{xi}$ and $a_{zi}$, explicitly constructing the actual driving surface current density on the microstrip metal, finding the resulting top or bottom boundary fields, and then utilizing operators to pull up or down through the structure layers, generating the electric and magnetic fields throughout in the process. The entire solution method uses the constraint that the vertical side walls of the device are perfect electric walls ( perfect metallic conductors), which can be shown to discretize the eigenvalues in the x – direction. These are the Fourier transform variables for the spectral domain, and an infinite set of them forms a complete set for the problem. Only a finite number of them are used, their maximum number being denoted by n.

## IV. Eigenvalues of LHM and RHM Devices at Millimeter Waves

To gain some idea of the general value of the propagation constant, an eigenvalue is sought for an ordinary medium substrate at the millimeter-wave frequency for a device with air above the substrate and right–handed medium (RHM/PPV, PPV) below the strip with $Re[\varepsilon(\omega)] = \varepsilon_r = 2.5$ and $Re[\mu(\omega)] = \mu_r = 2.3$, substrate thickness $h_s = 0.5$ mm, microstrip width w = 0.5 mm, air region thickness $h_a = 5.0$ mm, and vertical wall separation b = 5.0 mm. Also, $Im[\varepsilon(\omega)] = \varepsilon_i = 0$ and $Im[\mu(\omega)] = \mu_i = 0$ making the medium lossless (We also consider the microstrip metal lossless, although modifications for its loss can be made [46], as well as for medium loss in the substrate [46].). There are two $\gamma$ roots possible for even symmetry of the $J_z$ surface current component. They are mirror images of each other, with one corresponding to a z – directed wave and the other to a –z – directed wave propagating in the reverse longitudinal direction. For the fundamental mode, and we see that $\gamma = \alpha + j\beta = j\beta$, meaning that pure phase behavior occurs in the RHM/PPV structure. For a left–handed medium substrate, $\text{Re}[\varepsilon(\omega)] = -\varepsilon_r = -2.5$ and $\text{Re}[\mu(\omega)] = -\mu_r = -2.5$, with the geometric parameters being the same as for the RHM/PPV structure. Again, $\text{Im}[\varepsilon(\omega)] = -\varepsilon_i = 0$ and $\text{Im}[\mu(\omega)] = -\mu_i = 0$, meaning we will study the lossless case for the LHM/NPV structure. The LHM/NPV eigenvalues have been discussed at length elsewhere [35], and we only note here the contrast between the two types of eigenvalues, one being relatively simple (RHM/PPV), the other possessing regions of pure phase, evanescence, or multiple branch character (LHM/NPV). Because we are putting our attention deep into the millimeter wavelength regime, set frequency f = 80 GHz and find for the RHM/PPV structure $\bar{\beta} = \beta/k_0 = 2.200$. The dispersion diagram was produced with $n_x = n_z = 1$ and $n = 200$, although we have found solutions up to $n_x = n_z = 9$ and $n = 900$, the change in the numerical value being in the fourth decimal place.

For the LHM/NPV structure, two $\gamma$ solutions exist which have $\alpha = 0$, and $\bar{\beta} = \beta/k_0$ = 1.177647 and 1.78609 (quoted values for $n_x$ and $n_z$ = 1 and n = 200). One corresponds to a forward wave for low dispersive intrinsic LHMs ($\bar{\beta} = \beta/k_0 = 1.78609$) where the product of the integrated Poynting vector (net power through the cross–section) and phase vector in the z – direction is $\oint \mathbf{P}_z \bullet \beta \hat{\mathbf{z}} dA > 0$ (dA is the differential cross–sectional element) or equivalently $\mathbf{v}_{gl} \cdot \mathbf{v}_{pl} > 0$. The other solution is a backward wave for low dispersive intrinsic LHMs ($\bar{\beta} = \beta/k_0 = 1.177647$) where the product of the integrated Poynting vector and the phase vector in the z – direction is $\oint \mathbf{P}_z \bullet \beta \hat{\mathbf{z}} dA < 0$ or equivalently $\mathbf{v}_{gl} \cdot \mathbf{v}_{pl} < 0$. Here $\mathbf{v}_{gl}$ and $\mathbf{v}_{pl}$ are respectively the group and phase longitudinal velocities.

## V. Electromagnetic Field Distributions for Left – Handed Devices

In order to correctly perceive the intensity of the fields, $E^2 = \sum_{i=1}^{3} E_i^2 = E_t^2 + E_z^2$ and $H^2 = \sum_{i=1}^{3} H_i^2 = H_t^2 + H_z^2$ are calculated and plotted as E and H using a color linear scale. These field magnitudes are related to the overall energy content when pre-multiplied by the appropriate dispersive permittivity and permeability derivatives [1], [47], something that can be added to the model here by specifying additional microscopic physics of the LHM/NPV. This issue will be addressed in more detail elsewhere. What is important to realize here is that there is no conceptual difficulty in accomplishing that task. Care must be exercised in obtaining the linear color plot. Too small a grid results in a boxy appearance to the color distribution, making it very hard to interpret. Thus we seek on the order of $10^4$ grid points by partitioning each layer into 45 laminations and the vertical–to–vertical wall separation into 90 laminations, giving a 8372 grid points total from the spectral domain code. (Calculations have been done with as many as $2.5 \times 10^4$ grid points.) A Fortner algorithm is used to produce the finished color plots.

## A. RHM/PPV Comparison Structure

Once we have identified the $\bar{\beta}$ value and know the point in the diagram about which we wish to operate, we can proceed on to making a field distribution plot. Figures 2(a) and (b) respectively show, for the comparison RHM/PPV structure, the E and H magnitude plots for $\bar{\beta} = \beta/k_0 = 2.200$, with unscaled linear arrow plots of electric vector $\mathbf{E}_t$ and magnetic vector $\mathbf{H}_t$ overlaid on them. [Surface current coefficients are $a_{x1} = (0, -0.0637503)$ and $a_{z1} = (1, 0)$.] The arrow overlays allow us to assess the actual cross–sectional magnitudes of the fields locally, as well as their directions, whereas, the color distribution allows us to see continuously the entire magnitude of the fields. Field arrows are created in a $N \times M$ grid of points, $N \leq M$, usually $N = M - 1$, with $M = 22$ here. A number of basic features are seen in the plots. Variation of the magnitude has several periods in the x - direction associated with the effective wavelength in the RHM/PPV substrate. The electric field arrows emanating from the microstrip metal located about the device center line x = 0 mm, y = 0.5 mm, are directed out of the metal, consistent with a single charge residing on it, although it is in general nonuniformly distributed as expected for a conventional structure such as this. Magnetic field arrows circulate about the metal strip in one direction [Fig. 2(b)], breaking up into extra fine structure below the interface, but those near the metal strip continue into the substrate just under the strip in regions of extremely intense fields with singularities occurring based upon the edge condition. Even in the highly singular region near the metal strip which possesses delta function charge distributions near its edges, the discontinuity in the transverse $\mathbf{H}_t$ field is related to the surface current $\mathbf{J}$ to within 10 % or better by the cross – product $\mathbf{H}_t \times \hat{\mathbf{n}}$ where $\hat{\mathbf{n}} = \hat{y}$ = normal to the interface. Outside of the metal strip interface region, field arrows crossing the interface obey the necessary boundary conditions to within a few percent to small fractions of a percent (these observations also hold for the LHM substrate case).

### B. LHM/NPV Structure

At f = 80 GHz for the LHM/NPV substrate case, Figs. 3 (a) and (b) respectively show the electric E and magnetic H magnitude distributions for the lower root, backward wave LHM/NPV solution, with unscaled linear arrow plots of electric vector $\mathbf{E}_t$ and magnetic vector $\mathbf{H}_t$ overlaid on them. [Surface current coefficients are $a_{x1} = (0, -0.128419)$ and $a_{z1} = (1, 0)$.] We see from these plots that the electric field E resides both around the guiding strip as well as under it in the LHM/NPV substrate. Magnetic field H is more localized near the strip metal. The upper forward wave LHM/NPV solution shown in Figs. 4(a) and (b) demonstrates a more complicated distribution, with much more of it under the strip inside the LHM/NPV. [Surface current coefficients are $a_{x1} = (0, -0.142740)$ and $a_{z1} = (1, 0)$.] A number of striking differences are noted by comparing these results to the ordinary substrate medium case in Fig. 2 (RHM/PPV structure). Firstly, the E and H intensity distributions differ in appearance significantly for the LHM/NPV substrate structure compared to the relatively simple field pattern of the RHM/NPV structure. Secondly, in Fig. 3(a), the electric field arrows $\mathbf{E}_t$ do not point into (or out of) the metal strip but point roughly (this interpretation is modified by examining the field line distribution patterns to be discussed below in reference to Fig. 5 for the LHM/NPV structure) in one direction above and below the strip, indicating that this branch still has a single charge (as was found for the RHM/PPV case in Fig. 3) due to the reversed effect of the displacement electric field continuity condition normal to the interface. Thirdly, in Fig. 3(a), electric field $\mathbf{E}_t$ arrows, away from the metal strip near the interface, point in opposite directions as they cross the interface in terms of their normal components. Fourthly, in Fig. 3(b), magnetic field $\mathbf{H}_t$ arrows circulate around the strip in one direction above the interface and in the opposite direction below it, and the magnetic field arrows $\mathbf{H}_t$ when crossing the interface point in opposite directions in terms of their y – components. Fifthly, in Fig. 4(a), the electric field arrows $\mathbf{E}_t$ point roughly (again this interpretation is modified later by the field line distribution patterns to be presented below in Fig. 5) into (or out of) the metal guiding strip indicating a situation only possible if an infinitesimal dipolar charge arrangement exists in the vertical sense [35].

Of course, in addition to these noted differences between RHM/PPV and LHM/NPV substrates on guided wave behavior, the two solutions have significantly different field line distributions patterns as seen in Fig. 5 for the LHM/NPV structure. This is clear from the nearly circular circulating magnetic lines (first three) around the strip for the higher β value forward wave [Fig. 5(b)] compared to the broadly extended magnetic field lines for the lower β value backward wave [Fig. 5(a)] (This is evident by looking at the region above the substrate.). The electric field lines exhibit a dense pattern near the strip for the higher β value somewhat contained in a "shell," compared to that of the lower β value mini – "shell" which barely shows this pattern emerging (examine the region just above the metal strip in the air zone). For the lower β value [Fig. 5(a)], there is positive charge on the bottom half of the strip in the LHM/NPV (field lines enter into the strip) as well as on the ends (field lines exit from the last quarter length of the total strip length on either side of the strip) of the top part of the strip.  But the mini – "shell" has its $\mathbf{E}_t$ field lines emanating from the inner edges of the positive top charge at the boundary between this positive charge and an inside region of the top strip which is negatively charged.  Upper β solution has its "shell" extending about 20 % beyond the strip width, with the $\mathbf{E}_t$ field lines emanating from the "shell" surface with the lines terminating nearest the strip center, originating near the intersection of the "shell" surface and the interface.  Outside of the "shell" in both Figs. 5(a) and (b), the electric field lines $\mathbf{E}_t$ revert back to the simpler case of them seemingly to arise from a single uniform charge on both sides of the strip.  Because the "shell" is not seen at considerably lower frequencies, this seems to imply that at higher frequencies the structure is trying to become more like an ordinary media layered device.  Finally, we note that both the electric $\mathbf{E}_t$ and magnetic $\mathbf{H}_t$ field line patterns are much more intricate for the higher β value in comparison to the lower β value beneath the interface in the LHM/NPV substrate.

To better visualize the directions of the fields throughout the device area, unconstrained by the need to follow selected field lines, but to retain those useful features in such field line plots, we have developed scaled arrow plots which lift the tiny magnitude field values from the background.  One of the simplest methods to lift the fields in the background from numerical obscurity to visibility, is to perform a log scaling, such as,

$$E_i = \begin{cases} \dfrac{E_i}{E_{i,\min}} \log_{10}\left(\left|\dfrac{E_i}{E_{i,\min}}\right|\right) & ; |E_i| > E_{i,\min} \\ 0 & |E_i| \leq E_{i,\min} \end{cases} ; H_i = \begin{cases} \dfrac{H_i}{H_{i,\min}} \log_{10}\left(\left|\dfrac{H_i}{H_{i,\min}}\right|\right) & ; |H_i| > H_{i,\min} \\ 0 & |H_i| \leq H_{i,\min} \end{cases} \quad (2)$$

where $E_{i,\min}$ and $H_{i,\min}$ are positive magnitude cutoffs, and the argument presented to the $\log_{10}$ operator is always positive whereas the sign of the field component is preserved in the prefactor. Problem with this scaling approach, though, is that it can damage the angle between the x - and y - components of the fields. Therefore, we have instead created an inverse trigonometric method,

$$E_i = \dfrac{E_i}{E_t}\left[\left|\tan^{-1}\left(\dfrac{E_i}{E_{av}}\right)\right| + 0.75\right] ; H_i = \dfrac{H_i}{H_t}\left[\left|\tan^{-1}\left(\dfrac{H_i}{H_{av}}\right)\right| + 0.75\right] \quad (3)$$

Here

$$E_t^2 = E_x^2 + E_y^2 \quad ; \quad H_t^2 = H_x^2 + H_y^2 \quad (4)$$

Ratio factor on the left is is merely the cosine of the field angular offset from the x – axis, that is, $\cos\theta_E = E_i/E_t$ and $\cos\theta_H = H_i/H_t$. Figures 6(a) and (b) respectively plot the electric and magnetic field distributions for the lower $\beta$ LHM/NPV solution, with $E_{av} = 100$ V/m and $H_{av} = 0.01$ amps/m.

Lastly, in Fig. 7 is provided the electric field E magnitude distribution for the LHM/NPV structure, overlaid with electric field lines $\mathbf{E}_t$ for the upper eigenvalue at 80

GHz. This figure combines some of the information from Fig. 4 (a) and Fig. 5 (b), in such a way to assist visualization and understanding of the field behavior, allowing one to see the directional information at a glance while being able to assess the strength of the field.

## VI. Conclusion

In conclusion, we have shown completely new field distributions in a guided wave microstrip – like device structure containing left-handed material (LHM), otherwise referred to also as negative phase velocity material (NPV or NPVM) in the power/phase sense. The results are valid for intrinsic LH crystalline materials, LH metamaterials, or LH heterostructure or layered crystalline materials. These field distributions show new physics, and although that is what has drawn our attention, such structures which are compatible with integrated circuit and solid state technology, may open up many future possibilities. Unusual field distributions based upon new physics suggest the chance of completely new devices for future electronics besides the amendment of present devices which act as control components, active devices, and passive transmission structures. New devices could include millimeter-wave couplers, filters, phase shifters, isolators, and circulators, to mention a few.

# References

1. V. G. Veselago, in Advances in Electromagnetics of Complex Media & Metamaterials, 19, Kluwer, 2002.

2. P. F. Loschialpo, D. L. Smith, D. W. Forester & F. J. Rachford, Physical Rev. E (Rapid Commun.) **67**, 025602 (2003).

3. F. J. Rachford, D. L. Smith, P. F. Loschialpo, & D. W. Forester, Physical Rev. E **66**, 036613 (2002).

4. W. T. Lu, J. B. Sokoloff & S. Sridhar, Cornell Univ. Archive, arXiv: cond-mat/0306043v1, June 2 (2003).

5. A. Lakhtakia & J. A. Sherwin, Intern. J. Infrared Millimeter Waves **24**, 19 (2003).

6. A. Lakhtakia, Optics Express **11**, 716 (2003).

7. R.B. Greegor, C.G. Parazzoli, K. Li & M. H. Tanielian, Appl. Physics Letts. **82**, 2356 (2003).

8. S. Foteinpoulou, E. N. Economou & C. M. Soukoulis, Phys. Rev. Letts. **90**, 107402 (2003).

9. D. R. Smith & Schurig, Phys. Rev. Letts. **90**, 077405 (2003).

10. A. A. Houck, J. B. Brock, & I. L. Chuang, Phys. Rev. Letts, 90, 137401 (2003).

11. C. G. Parazzoli, R. B. Greegor, K. Li, B. E. C. Koltenbah & M. H. Tanielian, Phys. Rev. Letts. **90**, 107401 (2003).

12. J. Pacheco, Jr., T. M. Grzegorczyk, B.-I. Wu, Y. Zhang & J. A. Kong, Phys. Rev. Letts. **89**, 257401 (2002).

13. A. Lakhtakia, M. W. McCall & W. S. Weiglhofer, Int. J. Electron. Commun. **56**, 407 (2002).

14. R. Ruppin, J. Phys.: Condens. Matter **13**, 1811 (2001).

15. G. W. 't Hooft, Phys. Rev. Letts. **87**, 249701 (2001).

16. J. Pendry, Phys. Rev. Letts. **87**, 249702 (2001).

17. J. M. Williams, Phys. Rev. Letts **87**, 249703 (2001).

18. J. Pendry, Phys. Rev. Letts. **87**, 249704 (2001).

19. G. Dewar, Int. J. Mod. Physics **15**, 3258 (2001).

20. Z. M. Zhang & C. J. Fu, Appli. Phys. Letts. **80**, 1097 (2002).

21. A. Lakhatkia, Intern. J. Infrared Millimeter Waves **23**, 339 (2002).

22. E. Shamonina, V. A. Kalinin, K. H. Ringhofer & L. Solymar, Electr. Letts. **37**, 1243 (2001).

23. J. Paul, C. Christopoulos & D. W. P. Thomas, Electr. Letts. **37**, 912 (2001).

24. T. Weiland, R. Schulmann, R. B. Greegor, C. G. Parazzoli, A. M. Vetter, D. R. Smith, D. C. Vier & S. Schultz, J. Appli. Physics **90**, 5419 (2001).

25. C. Caloz, C. – C. Chang & T. Itoh, J. Appli. Physics **90**, 5483 (2001).

26. A. Grbic & G. V. Eleftheriades, IEEE Microwave Wireless Letts. **13**, 155 (2003).

27. G. V. Eleftheriades, A. K. Iyer & P. C. Kremer, IEEE Trans. Microwave Th. Tech. **50**, 2702 (2002).

28. A. Grbic & G. V. Eleftheriades, J. Appli. Physics **92**, 5930 (2002).

29. A. Alu & N. Engheta, Microwave & Optical Tech. Letts. **35**, 460 (2002).

30. N. Engheta, IEEE Antennas & Wireless Propag. Letts. **1**, 10 (2002).

31. C. Caloz, A. Sanada, L. Liu & T. Itoh, IEEE Microwave Th. Tech. Symp. Dig., 317 (2003).

32. I. H. Lin, C. Caloz, & T. Itoh, IEEE Microwave Th. Tech. Symp. Dig., 325 (2003).

33. H. Okabe, C. Caloz, & T. Itoh, IEEE Microwave Th. Tech. Symp. Dig., 329 (2003).

34. N. Engheta, , in Advances in Electromagnetics of Complex Media & Metamaterials, 83, Kluwer, 2002.

35. C. M. Krowne, Cornell Univ. Archive, arXiv.org/abs/physics/0305004, May 5 (2003). C. M. Krowne, Phys. Rev. Letts. **92**, 053901 (2004).

36. C. M. Krowne & M. Daniel, IEEE Microwave Th. Tech. Symp. Dig., 309 (2003).

37. C. M. Krowne, Bull. Am. Phys. Soc. **48**, Pt. 1, 580 (2003).

38. C. M. Krowne, IEEE Trans. Microwave Th. Tech. **51**, 2269 (2003).

39. C. M. Krowne, Bull. Am. Phys. Soc. **49**, Pt. 2, 928 ( 2004).

40. C. M. Krowne & A. Lakhatkia, in Negative Index Metamaterials DARPA Workshop: "Field Contouring & Focusing: Prospects for LHM Devices Based on Fundamental Physics," Arlington, VA., May 13, 2003.

41. Y. Zhang, B. Fluegel and A. Mascarenhas, Phys. Rev. Lett. **91**, 157404 (2003).

42. C. M. Krowne, Bull. Am. Phys. Soc. **49**, Pt. 2, 928, ( 2004).

43. C. M. Krowne, Phys. Rev. Letts. **92**, to be publ. (2004).

44. A. Lakhatkia & C. M. Krowne, Optik **114**, 305 (2003).  Also on Cornell Univ. Archive, arXiv, physics/0308043, Aug. 11 (2003).

45. C. J. B. Pendry, A. J. Holden, D. J. Robbins and W. J. Stewart, IEEE Trans. Micro. Th. Tech. **47**, 2075 (1999).

46. C. M. Krowne, IEEE Trans. Microwave Th. Tech. **50**, 112 (2002).  In (69), the numerator should read 1+j, not 1.  In (70), factors 1, 2, 1 are associated with (61), (64) and (65).

47. L. D. Landau, E. M. Lifshitz and L. P. Pitaevskii, Electrodynamics of Continuous Media, Pergamon Press 1984 (reprinted 1993).

**Figure Legends**

1. Cross-section of a microstrip structure with (a) a substrate using right-handed material (RHM/PPV) or (b) a substrate using left-handed material (LHM/NPV).

2. For an ordinary RHM/PPV substrate at 80 GHz, we have: (a) Electric field distributions in a color plot for magnitude E with an overlaid plot for vector $\mathbf{E}_t$ in arrow form unscaled; (b) Magnetic field distributions in a color plot for magnitude H with an overlaid plot for vector $\mathbf{H}_t$ in arrow form unscaled.

3. For a LHM/NPV substrate at 80 GHz, we have for the lower eigenvalue β: (a) Electric field distributions in a color plot for magnitude E with an overlaid plot for vector $\mathbf{E}_t$ in arrow form unscaled; (b) Magnetic field distributions in a color plot for magnitude H with an overlaid plot for vector $\mathbf{H}_t$ in arrow form unscaled.

4. For a LHM/NPV substrate at 80 GHz, we have for the upper eigenvalue β: (a) Electric field distributions in a color plot for magnitude E with an overlaid plot for vector $\mathbf{E}_t$ in arrow form unscaled; (b) Magnetic field distributions in a color plot for magnitude H with an overlaid plot for vector $\mathbf{H}_t$ in arrow form unscaled.

5. Electromagnetic field line distribution plots showing electric $\mathbf{E}_t$ (blue solid line) and magnetic $\mathbf{H}_t$ (red dashed line) fields at 80 GHz for a LHM/NPV substrate in the fundamental mode for (a) the lower $\beta/k_0 = 1.1776$ and (b) the upper $\beta/k_0 = 1.7886$ eigenvalue solutions.

6. Arrow field distribution plots for the LHM/NPV structure at 80 GHz for (a) electric field $\mathbf{E}_t$ and (b) magnetic field $\mathbf{H}_t$ for the lower $\beta/k_0 = 1.1776$ eigenvalue solution using a trigonometric scaling method.

7. For a LHM/NPV substrate at 80 GHz, for the upper eigenvalue β, is shown the electric field distribution in a color plot for magnitude E with an overlaid line distribution plot for $\mathbf{E}_t$.

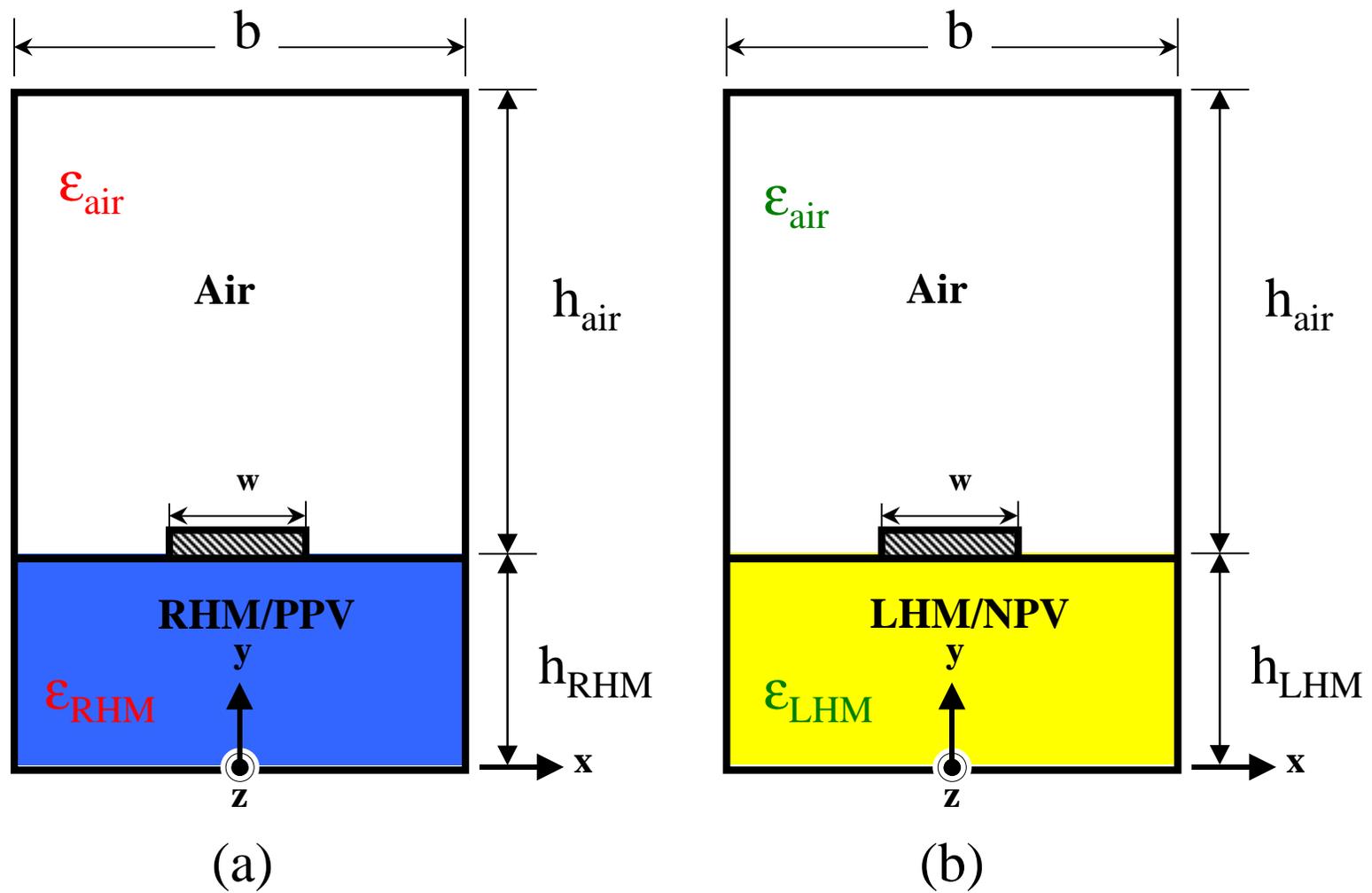

FIGURE 1

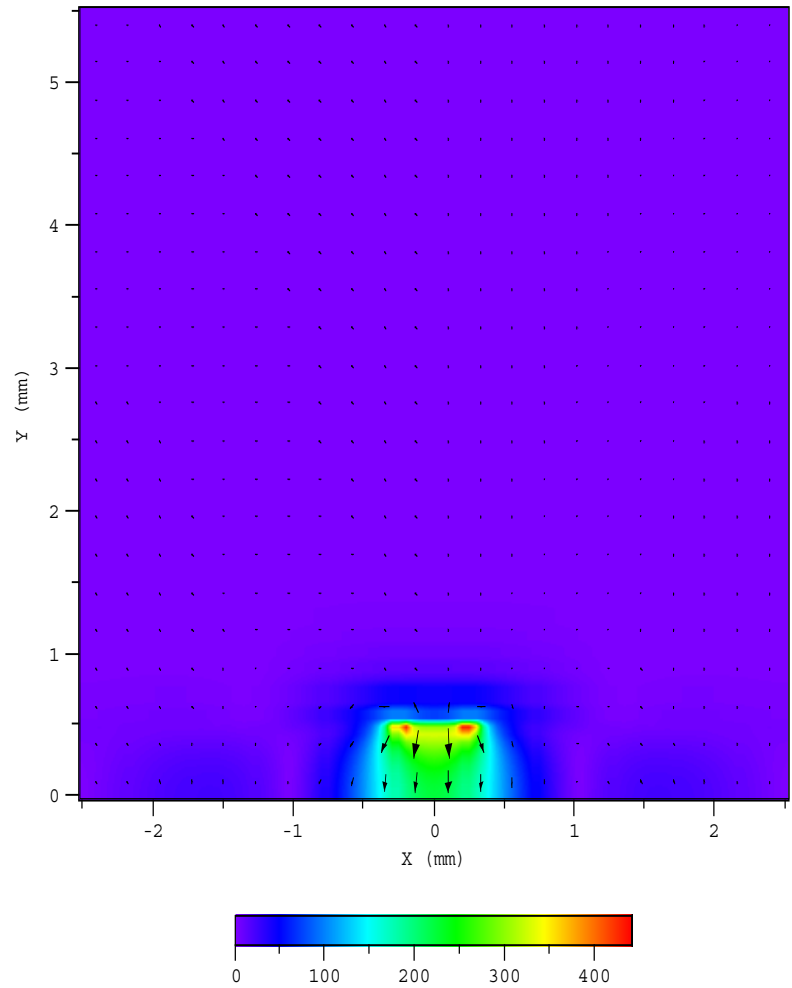

FIGURE 2 (a)

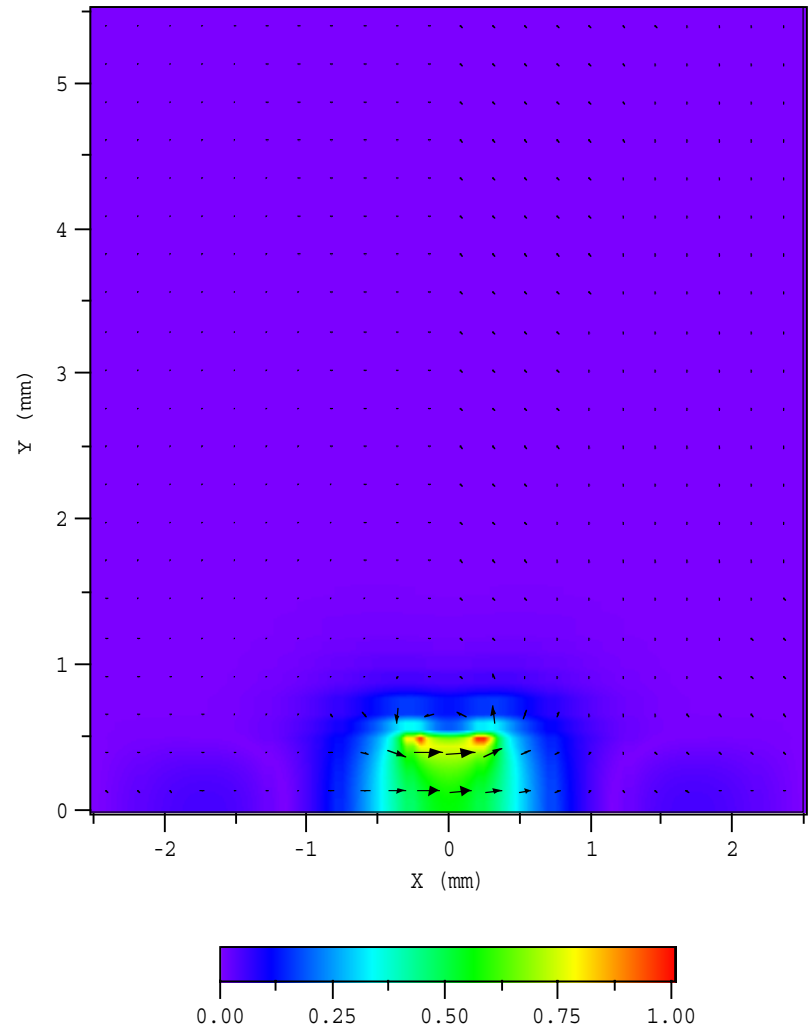

FIGURE 2 (b)

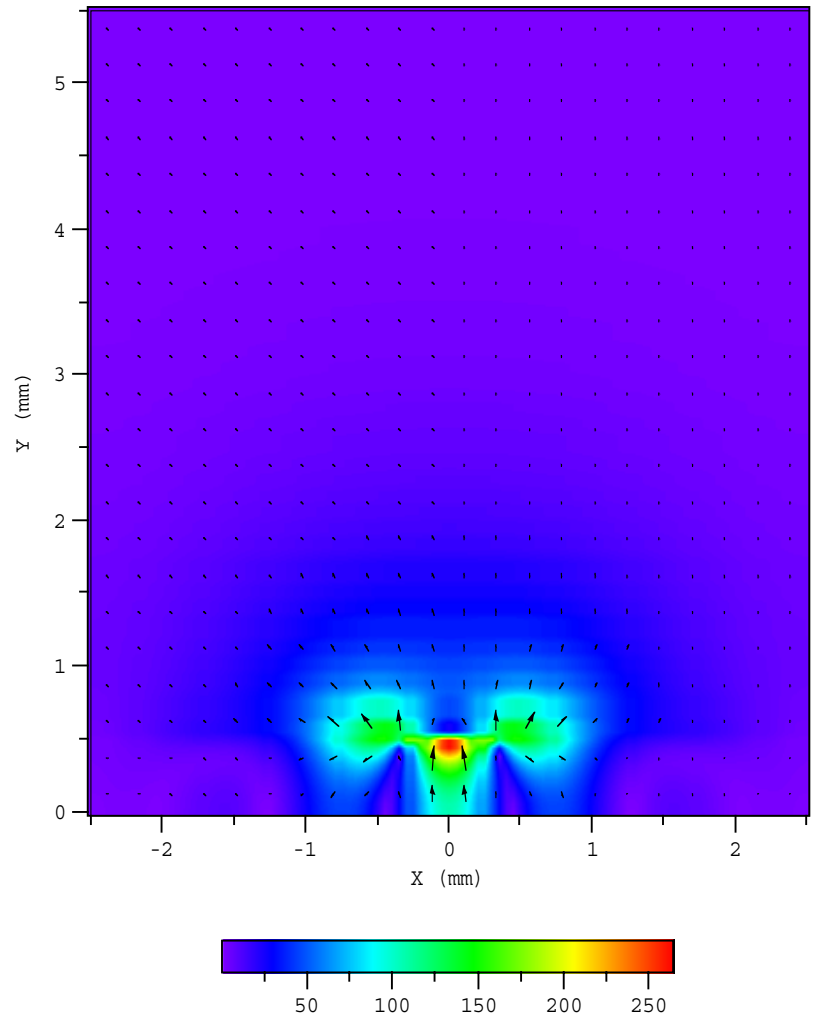

FIGURE 3 (a)

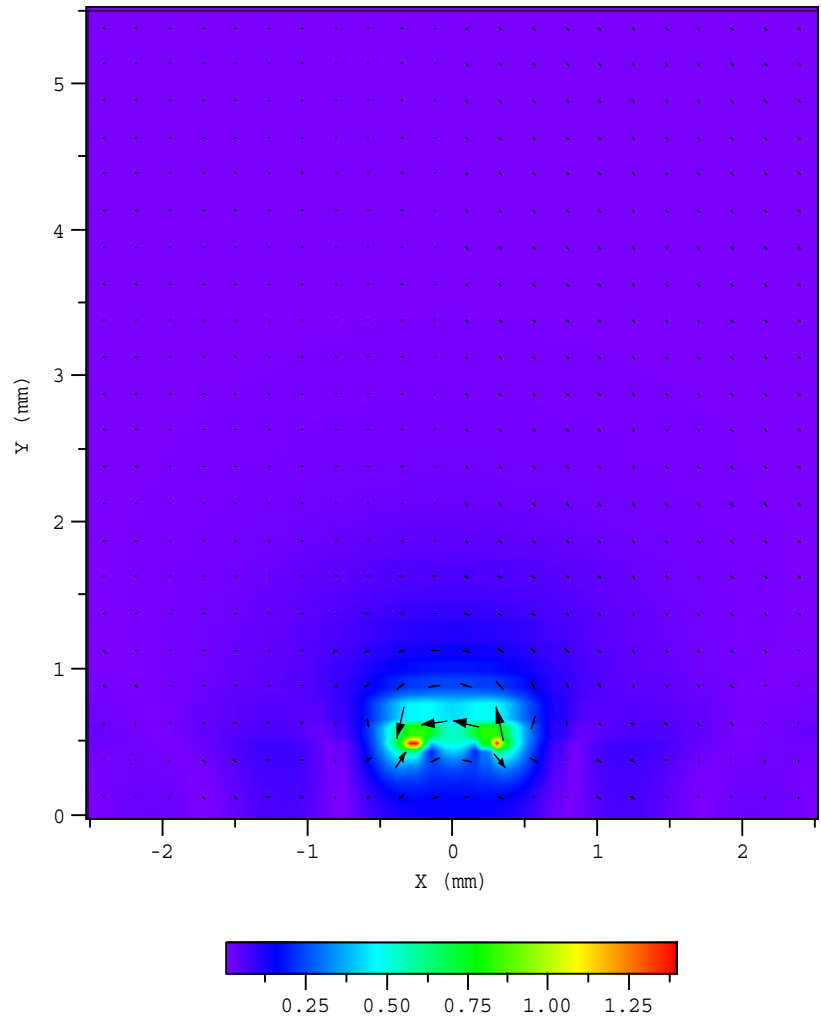

FIGURE 3 (b)

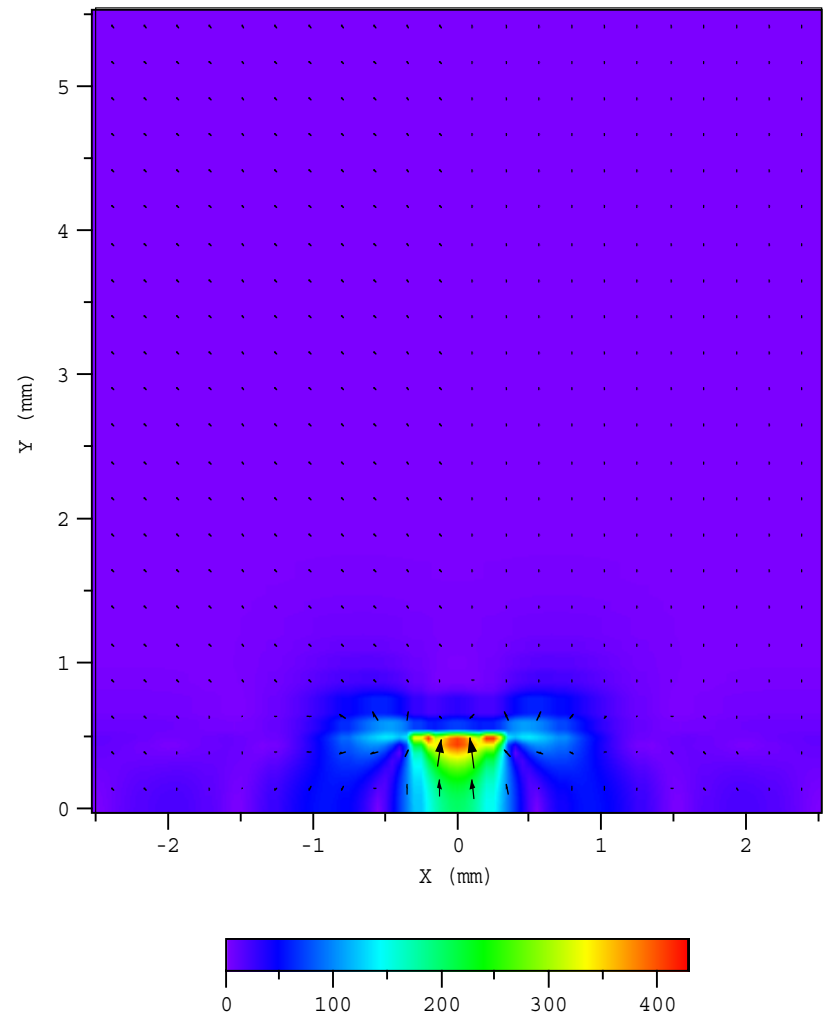

FIGURE 4 (a)

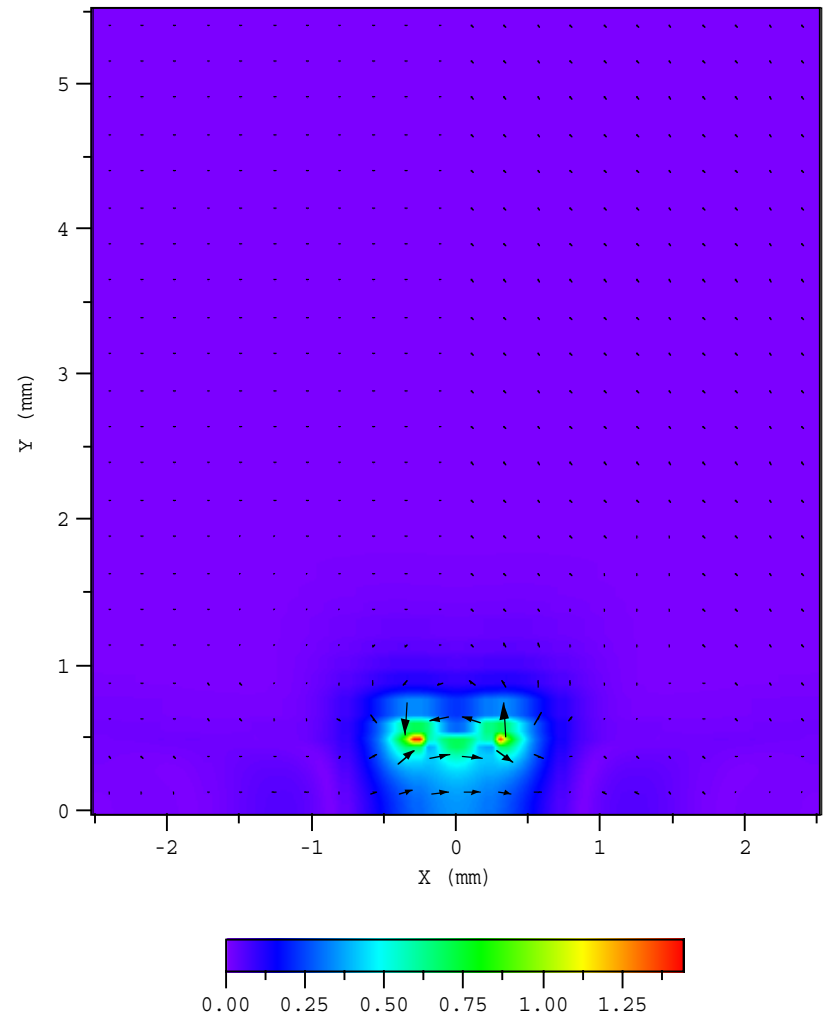

FIGURE 4 (b)

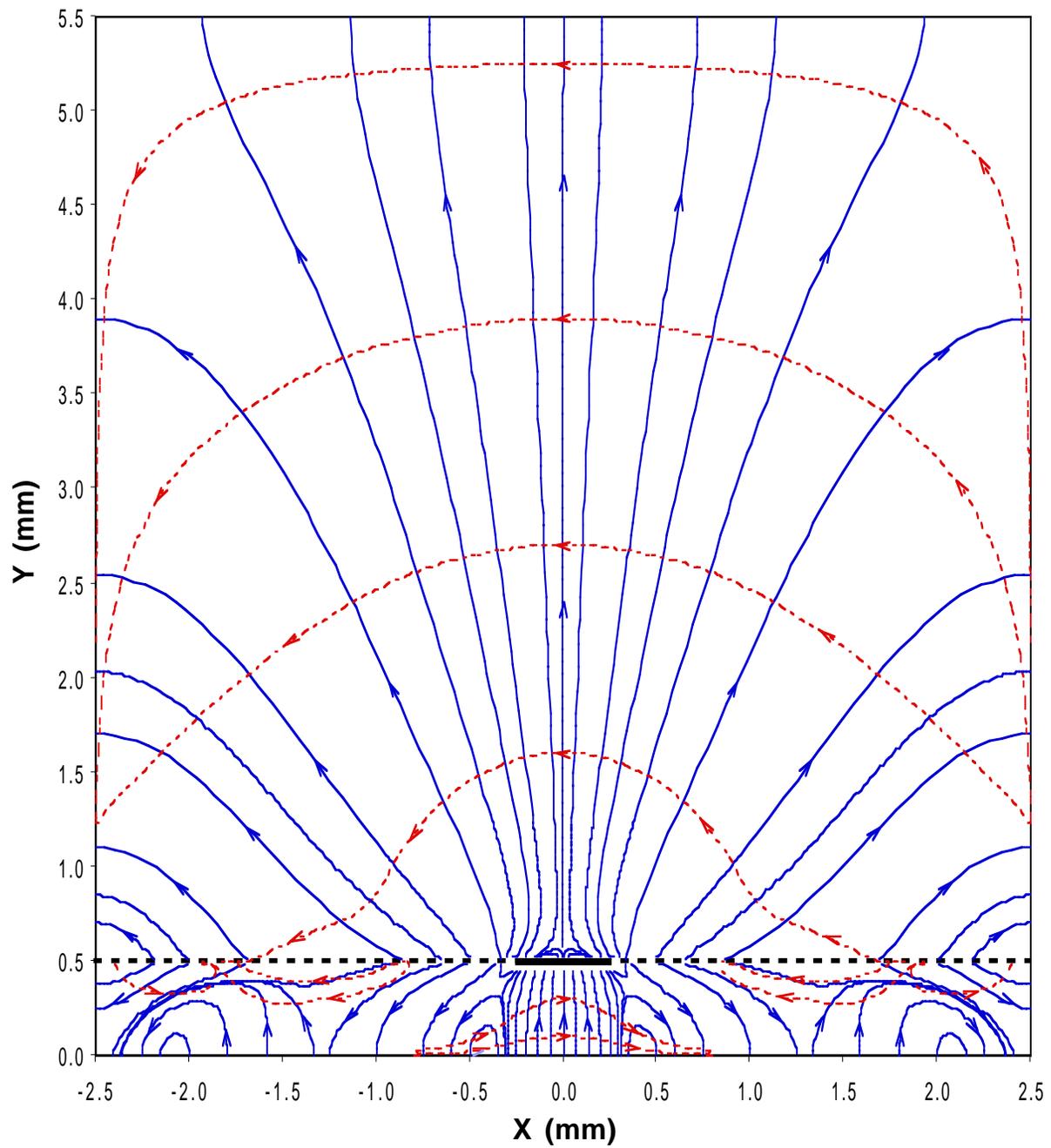

FIGURE 5 (a)

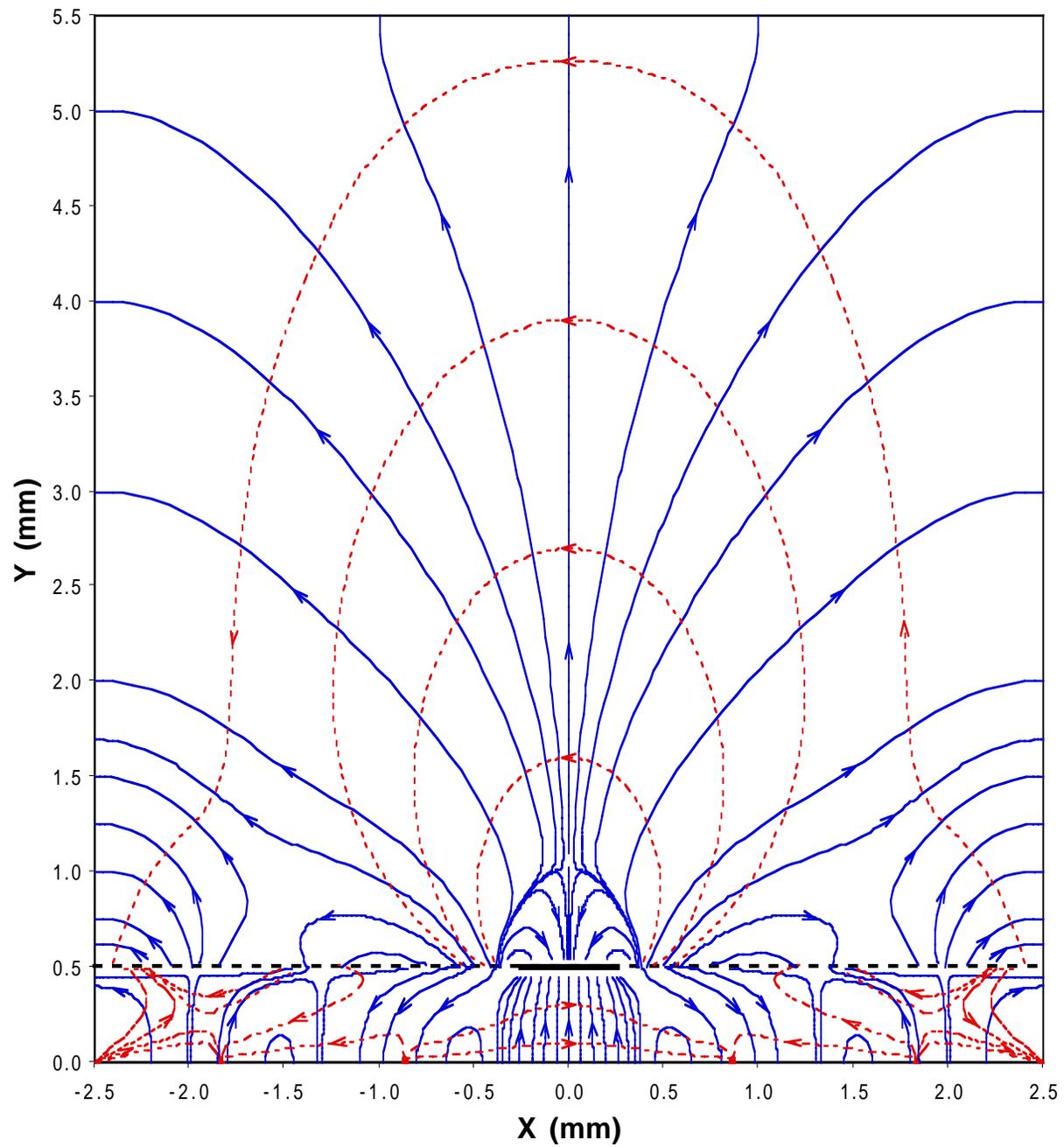

FIGURE 5 (b)

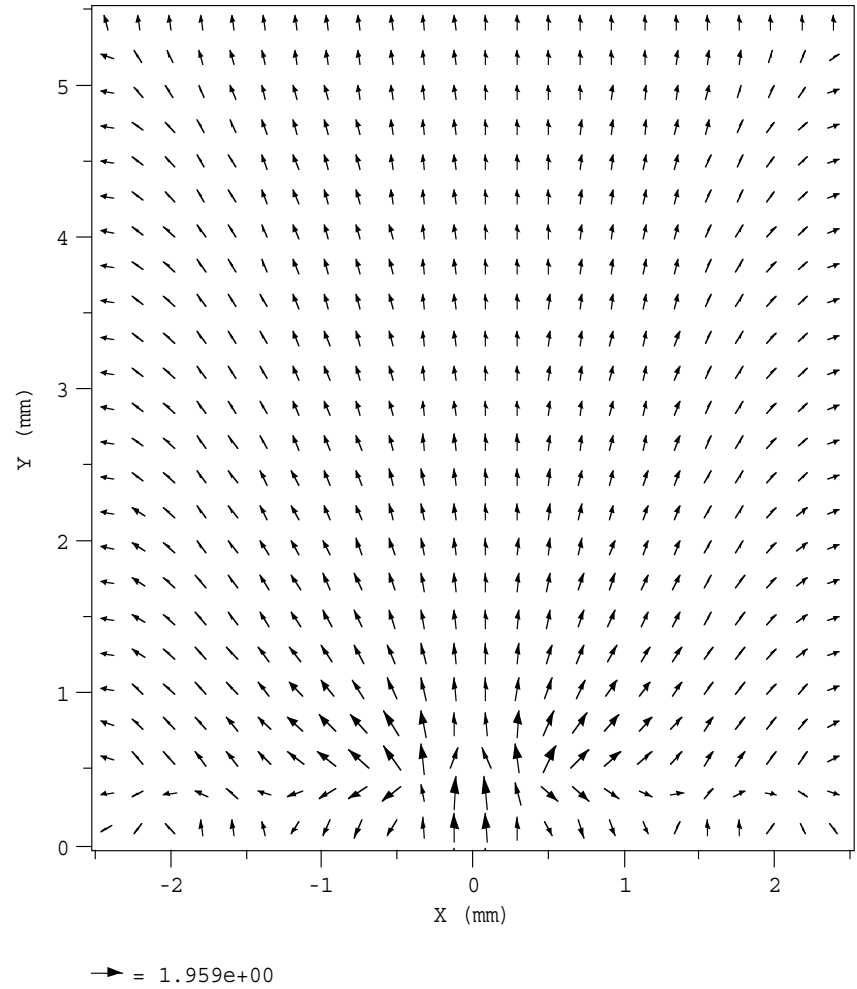

FIGURE 6 (a)

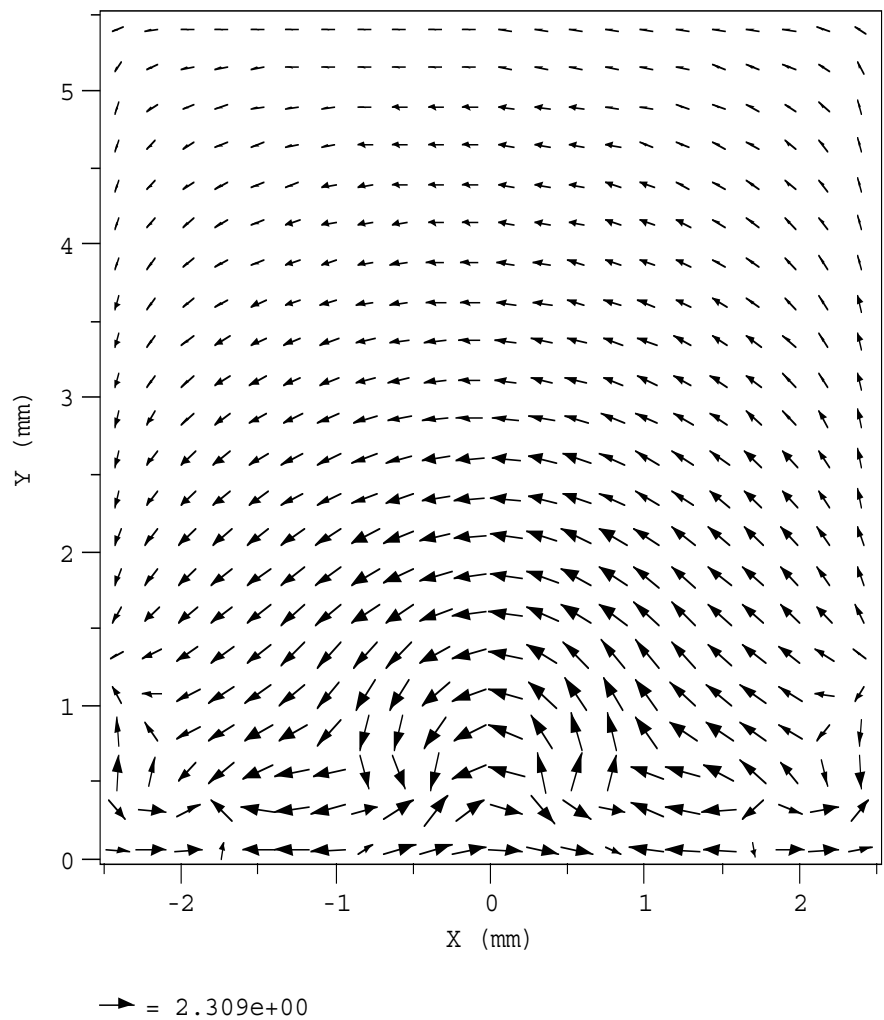

FIGURE 6 (b)

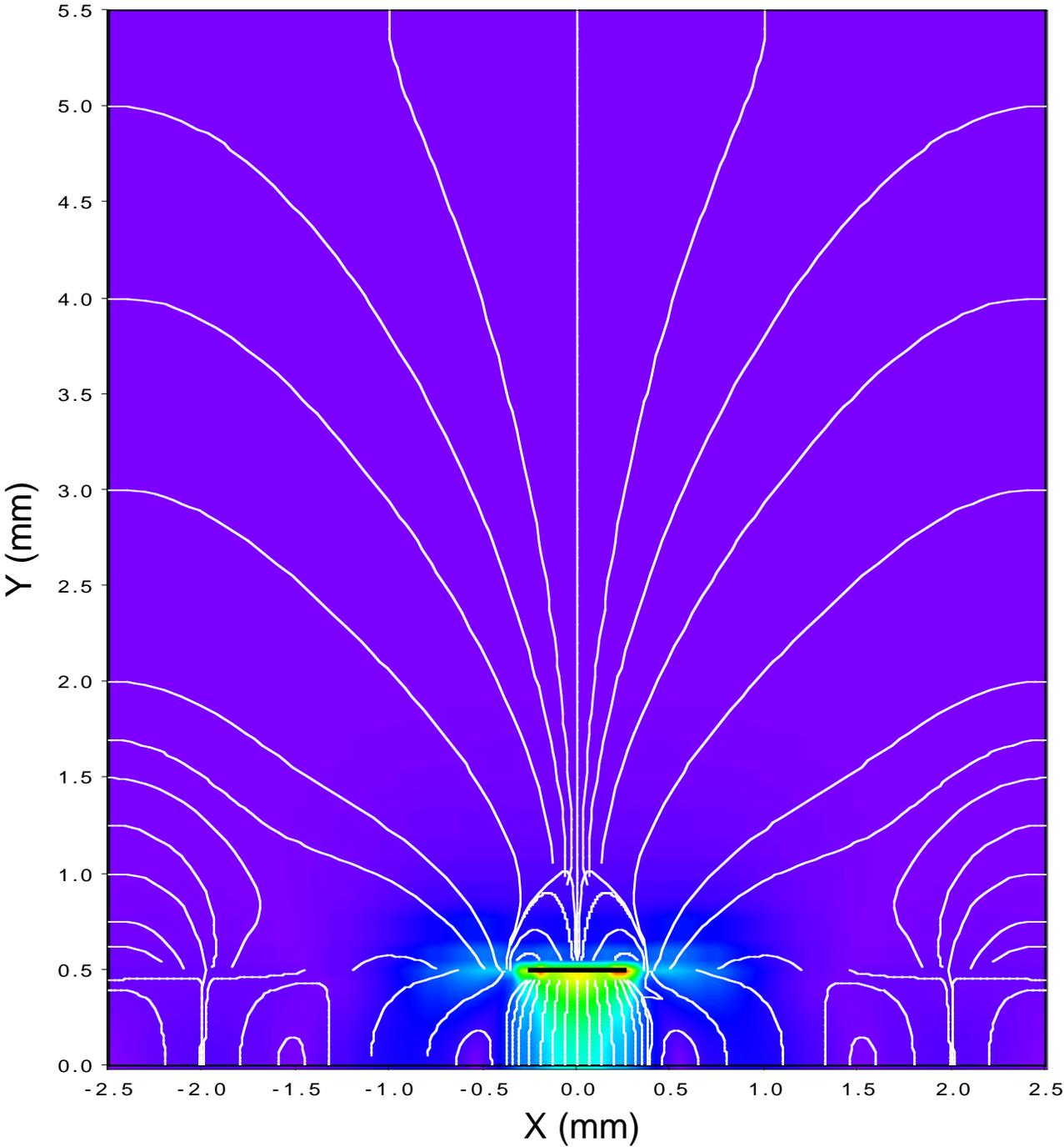

FIGURE 7